\documentclass[lettersize,journal]{IEEEtran}
\usepackage{amsmath,amsfonts}
\usepackage{algorithmic}
\usepackage{algorithm}
\usepackage{array}
\usepackage[caption=false,font=normalsize,labelfont=sf,textfont=sf]{subfig}
\usepackage{textcomp}
\usepackage{stfloats}
\usepackage{url}
\usepackage{verbatim}
\usepackage{graphicx}
\usepackage{cite}
\usepackage{hyperref}
\usepackage{multirow}
\usepackage{booktabs}
\usepackage{soul}
\usepackage{xcolor}

\begin{document}

\title{Unsupervised Training of a Dynamic Context-Aware Deep Denoising Framework for Low-Dose Fluoroscopic Imaging}
\author{Sun-Young Jeon, Sen Wang, Adam S. Wang, Garry E. Gold, and Jang-Hwan Choi, \IEEEmembership{Member, IEEE}
\thanks{This research was partly supported by the Technology Development Program of MSS [S3146559]; by the National Research Foundation of Korea (NRF-2022R1A2C1092072); and by the Korea Medical Device Development Fund grant funded by the Korea government (the Ministry of Science and ICT, the Ministry of Trade, Industry and Energy, the Ministry of Health \& Welfare, the Ministry of Food and Drug Safety) (Project Number: 1711174276, RS-2020-KD000016).

}
\thanks{S.-Y. Jeon and J.-H. Choi are with the Department of Artificial Intelligence Convergence, and J.-H. Choi is also with the Department of Computational Medicine, Graduate Program in System Health Science and Engineering, Ewha Womans University, Seoul, South Korea (e-mail: sunyounge\_@ewhain.net; choij@ewha.ac.kr).}
\thanks{S. Wang, A. S. Wang, and G. E. Gold are with the Department of Radiology, Stanford University, Stanford, California, USA (e-mail: senwang@stanford.edu; adamwang@stanford.edu).}}

\maketitle

\begin{abstract}
Fluoroscopy is critical for real-time X-ray visualization in medical imaging. However, low-dose images are compromised by noise, potentially affecting diagnostic accuracy. Noise reduction is crucial for maintaining image quality, especially given such challenges as motion artifacts and the limited availability of clean data in medical imaging. To address these issues, we propose an unsupervised training framework for dynamic context-aware denoising of fluoroscopy image sequences. First, we train the multi-scale recurrent attention U-Net (MSR2AU-Net) without requiring clean data to address the initial noise. Second, we incorporate a knowledge distillation-based uncorrelated noise suppression module and a recursive filtering-based correlated noise suppression module enhanced with motion compensation to further improve motion compensation and achieve superior denoising performance. Finally, we introduce a novel approach by combining these modules with a pixel-wise dynamic object motion cross-fusion matrix, designed to adapt to motion, and an edge-preserving loss for precise detail retention. To validate the proposed method, we conducted extensive numerical experiments on medical image datasets, including 3500 fluoroscopy images from dynamic phantoms (2,400 images for training, 1,100 for testing) and 350 clinical images from a spinal surgery patient. Moreover, we demonstrated the robustness of our approach across different imaging modalities by testing it on the publicly available 2016 Low Dose CT Grand Challenge dataset, using 4,800 images for training and 1,136 for testing. The results demonstrate that the proposed approach outperforms state-of-the-art unsupervised algorithms in both visual quality and quantitative evaluation while achieving comparable performance to well-established supervised learning methods across low-dose fluoroscopy and CT imaging. The related source code will be available at \href{https://github.com/sunyoungIT/UDCA-Net.git}{https://github.com/sunyoungIT/UDCA-Net.git}.
\end{abstract}


\section{Introduction}
\label{sec:introduction}
\begin{figure}[ht]
\centerline{\includegraphics[width=\linewidth]{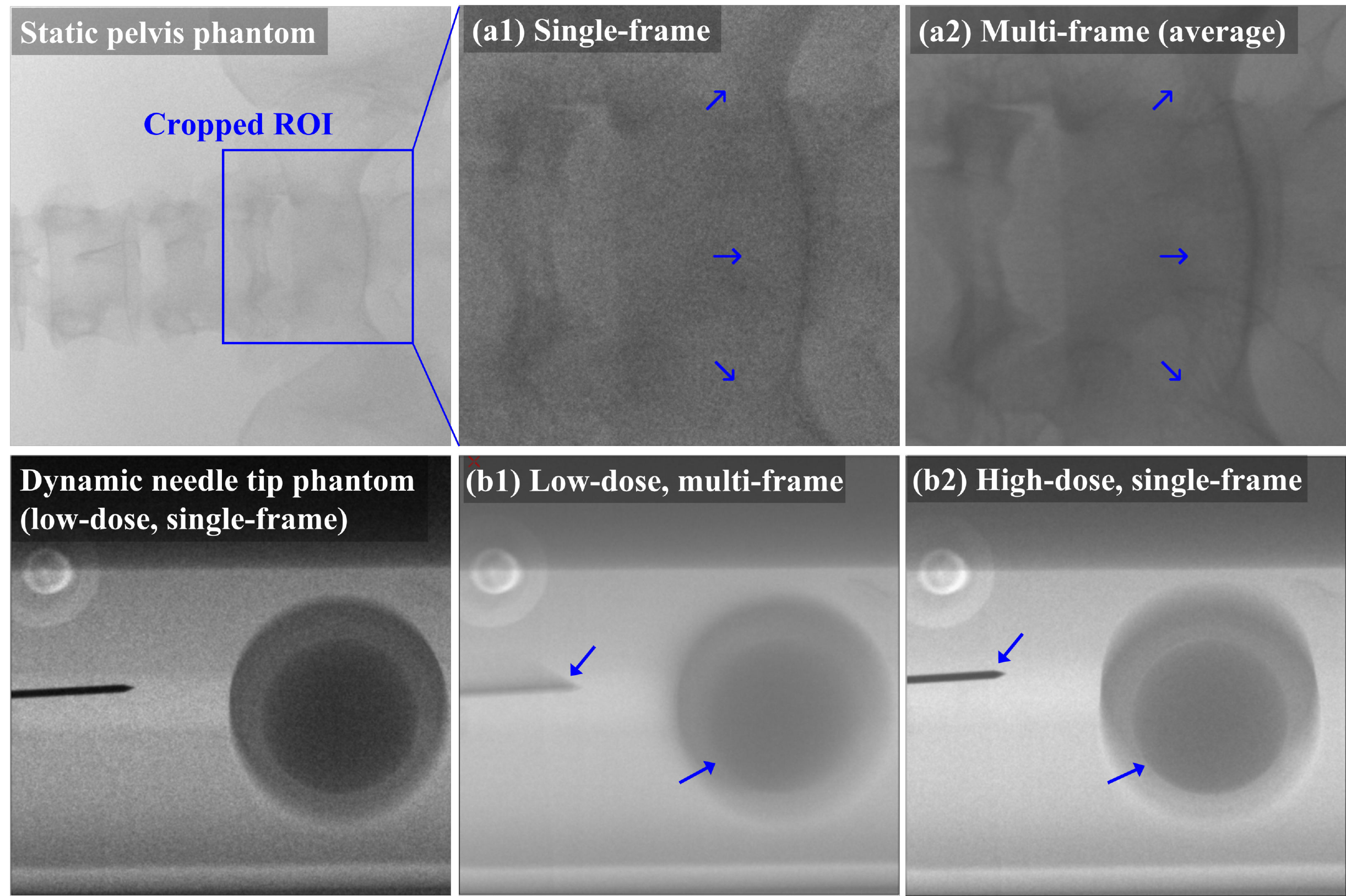}}
   \caption{Typical fluoroscopic images obtained from (a) a static pelvis phantom and (b) a dynamic needle tip with a spherical lesion phantom.}
\label{Fig1} 
\end{figure} 
\IEEEPARstart{F}{luoroscopy} is critical in medical imaging, enabling the real-time generation and visualization of X‑ray images. It is invaluable in understanding internal body structures and organs and imaging medical devices, such as catheters. Consequently, fluoroscopy is an indispensable tool used in various diagnostic examinations, including catheter insertion, vascular imaging, and monitoring during orthopedic surgery. However, the inherent ionizing radiation from X-rays presents considerable risks to patients and medical personnel~\cite{li2020sacnn}. Reducing the X-ray intensity is recommended to mitigate this risk. Adopting low-dose X-ray fluoroscopy is a standard practice in monitoring interventions. Although lower doses reduce radiation exposure, they can produce images with increased noise and artifacts~\cite{andreozzi2020novel}. Such imperfections can obscure vital details, potentially affecting clinical decision-making. Thus, efficient noise reduction techniques tailored to low-dose X-ray fluoroscopic images are necessary to maintain essential features, including medical instruments and vital anatomical structures.

Various restoration techniques have been introduced to mitigate noise-caused image degradation. Their efficiency is closely related to the suitability of the noise model. Many of these methods characterize the noise as either an additive spatially invariant Gaussian~\cite{cerciello2011advanced, luo2020ultra} or a signal-dependent Poisson~\cite{rodrigues2008denoising, cesarelli2013x}. However, these noise-specific assumptions may not universally apply to clinics equipped with various fluoroscopy devices. Thus, demand exists for denoising algorithms without a specific noise model. Because of the significant noise levels in fluoroscopic images~\cite{cesarelli2013x}, restoring signals obscured by noise in a single fluoroscopic frame is unattainable. For instance, as demonstrated in Fig.~\ref{Fig1}(a), bone tissue obscured in a single frame (a1) becomes evident when averaging multiple frames of fluoroscopic image sequences (a2). Therefore, denoising fluoroscopic images must be performed using multi-frame image sequences, not just a single image. 

Multi-frame processing algorithm using traditional temporal filters has been proposed~\cite{cesarelli2013x,tomic2012adaptive,wagner2015noise,cerciello2010noise,schoonenberg2005adaptive}. These traditional filter-based methods have successfully reduced noise and improved the signal-to-noise ratio (SNR) in low-dose fluoroscopic images while minimizing processing time. However, depending on the specific filter applied, the resulting filtered images may exhibit motion-induced blurring in the temporal direction. For instance, as depicted in Fig.~\ref{Fig1}(b), when applying a recursive filter to a multi-frame sequence image (b1), significant blurring is caused by motion that is not present in the single frame image (b2). To address motion-blurring problems, many researchers have developed methods to enhance the quality of sequential fluoroscopic images by diminishing noise artifacts~\cite{476114,wilson1999perception,nishiki2008method,jin2006wavelet,amiot2016spatio,hariharan2018photon,chan1993image}. For example, Amiot~\textit{et al.}~\cite{amiot2016spatio} applied a motion-compensated temporal filtering technique operating on multi-scale coefficients to image sequences, successfully capturing subtle features with low contrast while minimizing boundary blurring. However, when acquiring images in X-ray fluoroscopy sequences involving complex motion, motion artifacts, such as blurring and trailing, can occur~\cite{lee2017motion, sarno2019real, schoonenberg2005adaptive}. This problem underscores the need for proper motion compensation when handling multi-frame sequence images.
 
With the recent emergence of deep learning technology, significant advancements have been achieved in various image processing tasks, including image denoising~\cite{abdelhamed2020ntire,yan2023image, ye2021detail,choi2020statnet,marcos2022dilated,liu2022dfsne,kang2022denoising,chen2024lit,li2023dual,wu2022masked,wang2021x,chao2022dual,li2021incorporation,wu2023deep,chang2019two}, to enhance denoising performance. This development has yielded remarkable results, demonstrating the successful application of deep learning in image denoising. Furthermore, several convolutional neural network (CNN)-based methods have been proposed in the field of fluoroscopy images~\cite{luo2022edge, luo2020ultra, zhang2019hybrid, van2021real,wu2020combined,kim2022deep}. These approaches successfully reduce noise while retaining image details, displaying exceptional performance. However, most approaches for denoising fluoroscopy images employ supervised learning and rely on a substantial dataset of paired images. These paired images include noisy low- and high-dose X-ray images, which are essential to learn the mapping function between the two image types effectively. Obtaining such paired data in a clinical interventional setting presents significant challenges due to patient motion over fluoroscopy image sequences and the potential risks associated with increased ionizing radiation exposure. To address this problem, self-supervised or unsupervised learning-based methods have been proposed for medical image denoising, particularly in the context of CT imaging~\cite{bai2021probabilistic,kim2022deep,jing2022training,jeon2022mm,wagner2023noise2contrast,tang2019unpaired,yuan2020half2half,lagerwerf2020noise2filter,zhang2021noise2context,zhao2023dual,huang2021gan}. However, research on fluoroscopy image denoising remains limited, and studies in this field are notably scarce~\cite{juneja2024denoising}. This represents a significant gap in the application of advanced denoising techniques to fluoroscopy. Despite the limited number of studies on unsupervised and self-supervised methods for fluoroscopy noise reduction, a few notable examples do exist. For instance, Liu \textit{et al.}~\cite{liu2022stabilize} proposed a three-stage, self-supervised framework for denoising fluoroscopy videos, involving stabilization, mask-based RPCA decomposition, and spatiotemporal bilateral filtering. This approach underscores the potential of enhancing self-supervised learning techniques in fluoroscopy applications. Similarly, Sanderson\textit{et al.}~\cite{sanderson2024diffusion} introduced DDPM-X, a diffusion model-based denoising method for planar X-ray images, capable of handling Gaussian and Poisson noise without network modifications. However, Liu \textit{et al.}~\cite{liu2022stabilize} utilized the Self2Self~\cite{quan2020self2self}, making it challenging to specify the noise model, particularly in real-world scenarios where such models are crucial. Furthermore, preserving edge details is essential for accurate diagnostics, and Sanderson \textit{et al.}~\cite{sanderson2024diffusion} highlights the need for proper motion compensation methods for moving fluoroscopy sequences. These challenges underscore the key requirements for effective fluoroscopic image denoising algorithms, which include 1) not relying on a specific noise model, 2) the necessity of multi-frame sequence images, 3) the need for proper motion compensation in the temporal domain, and 4) the requirement for unsupervised learning due to patient movement. 

Given these requirements, we propose an unsupervised, two-step denoising framework that effectively utilizes multi-frame sequence images to meet these critical needs. Our approach specifically addresses motion compensation, multi-frame utilization, and unsupervised learning, while also enhancing edge preservation and improving overall image quality in fluoroscopy. In the first step, we pretrain the multi-scale recurrent attention U-Net (MSR2AU-Net) using a self-supervised approach that predicts the middle frame from a series of sequential fluoroscopic images. In the following step, we incorporate a correlated noise suppression module to manage denoising in pixel regions with significant inter-frame motion and an uncorrelated noise suppression module for areas with less motion. The correlated noise suppression module operates recursive filtering on sequence images that have undergone motion compensation. We adopt a knowledge distillation approach for the uncorrelated noise suppression module, using the pre-trained MSR2AU-Net as a teacher model to train a student denoising model. We aim to preserve image edges and ensure outstanding denoising outcomes by fusing these modules and tailoring them to the dynamic object motion on a pixel-by-pixel basis. We evaluate the denoising performance of the unsupervised two-step framework using dynamic phantom and clinical, in vivo datasets. We also used the publicly available CT dataset from the ``2016 NIH-AAPM-Mayo Clinic Low Dose CT Grand Challenge'' dataset~\cite{mayochallenge}. We compare it with the current state-of-the-art unsupervised and supervised learning methods. The experimental results indicate that the proposed framework effectively minimizes noise levels while maintaining the intricate details of microstructures. The primary contributions of this paper are summarized as follows:
\begin{itemize}
    \item An unsupervised dynamic context-aware denoising framework is proposed to mitigate both correlated and uncorrelated motion-induced quantum noise in low-dose fluoroscopy images. To the best of our knowledge, this is the first attempt at an unsupervised deep learning-based denoising approach specifically designed for dynamic low-dose X-ray fluoroscopy.

    \item Based on dynamic context extraction modules, motion compensation and recursive filtering modules are integrated into the network to enhance denoising capabilities and effectively address issues like motion blurring by considering temporal correlations between frames.
    
    \item In the edge-preserving extraction module, a pixel-wise cross-fusion matrix is designed to capture changes between the motion-corrected image and noise reduction, maintaining sharp edges and ensuring optimal denoising results. This module combines correlated and uncorrelated noise reduction techniques, adapting to the dynamic motion of the object on a pixel-by-pixel basis.

    \item To improve the network's ability to maintain sharpness and enhance perceptual quality, an improved loss function is proposed to maximize the extraction of high-frequency features by combining Wavelet and Fourier Transform techniques.

    \item The evaluations on clinical and dynamic-phantom fluoroscopy datasets, as well as low-dose CT, reveal that the proposed algorithm not only surpasses existing leading unsupervised algorithms but also matches the performance of supervised learning-based approaches. Additionally, our framework has been effectively applied to both low-dose CT and fluoroscopy, showcasing strong generalization across different medical imaging modalities.
\end{itemize}

\subsection{Related Works}
\subsubsection{Low-dose CT Denoising}
Various CNN-based methods, such as U-Net~\cite{ronneberger2015unet} and DnCNN~\cite{zhang2017beyond}, have been employed to reduce noise in low-dose CT (LDCT) images. Similarly, RED-CNN~\cite{7947200} utilizes a combination of ResNet and Autoencoder architectures to further improve denoising performance. These CNN-based approaches showcase the considerable potential of deep learning techniques in advancing medical imaging. However, these techniques often produce overly smooth images due to the mean squared error (MSE) loss. To address this, alternatives like perceptual and GAN losses have been introduced. For instance, exemplified in Yang et al.'s WGAN-VGG framework~\cite{8340157}, effectively address this smoothing issue, resulting in more realistic image details and improving overall image quality. Kim \textit{et al.}~\cite{kim2022wavelet} proposed a method combining pixel-wise losses for high objective quality with perceptual and wavelet losses to preserve fine details and edges, leveraging wavelet transform properties for enhanced image quality. In addition, CCN-CL~\cite{tang2022ccn} enhances denoising accuracy using a content-noise complementary learning strategy, attention mechanisms, and deformable convolutions. Transformer-based models like CTformer~\cite{wang2023ctformer} introduce pure transformer approaches to LDCT denoising, outperforming traditional methods. StruNet~\cite{ma2023strunet} presents a novel Swin transformer-based residual U-shape network, showing promise in handling diverse noise artifacts across various imaging modalities. However, these networks generally rely on supervised learning with paired LDCT and normal-dose CT datasets, a method challenging for fluoroscopy imaging, which involves capturing moving objects. Recently, self-supervised models that only require noisy natural images (e.g., BM3D~\cite{dabov2007image}, NLM~\cite{li2014adaptive}, Noisier2Noise~\cite{moran2020noisier2noise}, Noise2Void~\cite{krull2019noise2void}, and Noisy-As-Clean~\cite{xu2020noisy}) have been applied for denoising tasks in both LDCT and fluoroscopy imaging. These self-supervised approaches have shown promising results, demonstrating their potential to effectively reduce noise without the need for clean reference images, making them valuable tools for improving image quality in challenging medical imaging scenarios. Nonetheless, these methods rely on specific assumptions about noise characteristics in the images, and therefore, their application in fluoroscopic imaging might result in suboptimal performance. This highlights the importance of exploring unsupervised learning techniques, which can provide more flexible and generalized solutions for noise reduction without relying on such assumptions. MM-Net~\cite{jeon2022mm}, a framework based on unsupervised learning, incorporates multi-mask patching with high-frequency components, allowing it to surpass existing unsupervised algorithms in both qualitative and quantitative assessments across diverse clinical and animal data domains. DenoisingGAN~\cite{kim2024unsupervised} leverages unpaired datasets and integrates a CycleGAN variation with a memory-efficient architecture, demonstrating superior performance in both objective and perceptual quality. These unsupervised learning approaches offer significant advantages by not relying on paired NDCT and LDCT images, making them more adaptable and easier to implement in various clinical settings. However, their clinical application remains limited due to the current limitations in denoising performance.
 
\subsubsection{Low-dose X-ray Fluoroscopy Denoising}
While there are relatively fewer deep learning-based denoising algorithms for fluoroscopy compared to LDCT, recent advancements have been made in this area. For instance, Zhang et al. \cite{zhang2019hybrid} proposed a hybrid three-dimensional (3D)/two-dimensional (2D)-based deep CNN framework for X-ray angiography using a stack of consecutive frames as input. Unlike single-frame approaches, which struggle with motion artifacts due to their limited temporal context, this method harnesses both spatial and temporal information from multiple frames. This comprehensive approach not only significantly enhances image quality but also effectively reduces motion artifacts and preserves fine details, demonstrating a marked improvement over traditional single-frame denoising techniques. In addition, Wu et al. \cite{wu2020combined} introduced the multi-channel DnCNN (MCDnCNN) based on DenseNet \cite{huang2017densely} and DnCNN \cite{zhang2017beyond}, demonstrating enhanced denoising capabilities for fluoroscopic sequence images. Van Veen et al. \cite{van2021real} showcased clinically viable video denoising results for fluoroscopic imaging using a network based on FastDVDNet~\cite{tassano2020fastdvdnet}, which incorporates multiple adjacent frames. This study integrates motion estimation directly into the network architecture, significantly enhancing runtime efficiency while maintaining or improving video quality and reducing radiation dose. Moreover, Luo et al. proposed UDDN~\cite{luo2020ultra}, utilizing dense connections for information reuse in CNN frameworks, thereby enhancing denoising performance. They further improved this approach in EEDN~\cite{luo2022edge} by incorporating an edge-enhancement module, which significantly sharpens edges and fine details, resulting in clearer and more defined images. However, all these algorithms are based on supervised learning, making their application to clinical fluoroscopy imaging quite challenging.

\section{METHODS}
\label{sec:method}
\begin{figure}[ht!]
\centerline{\includegraphics[width=\linewidth]{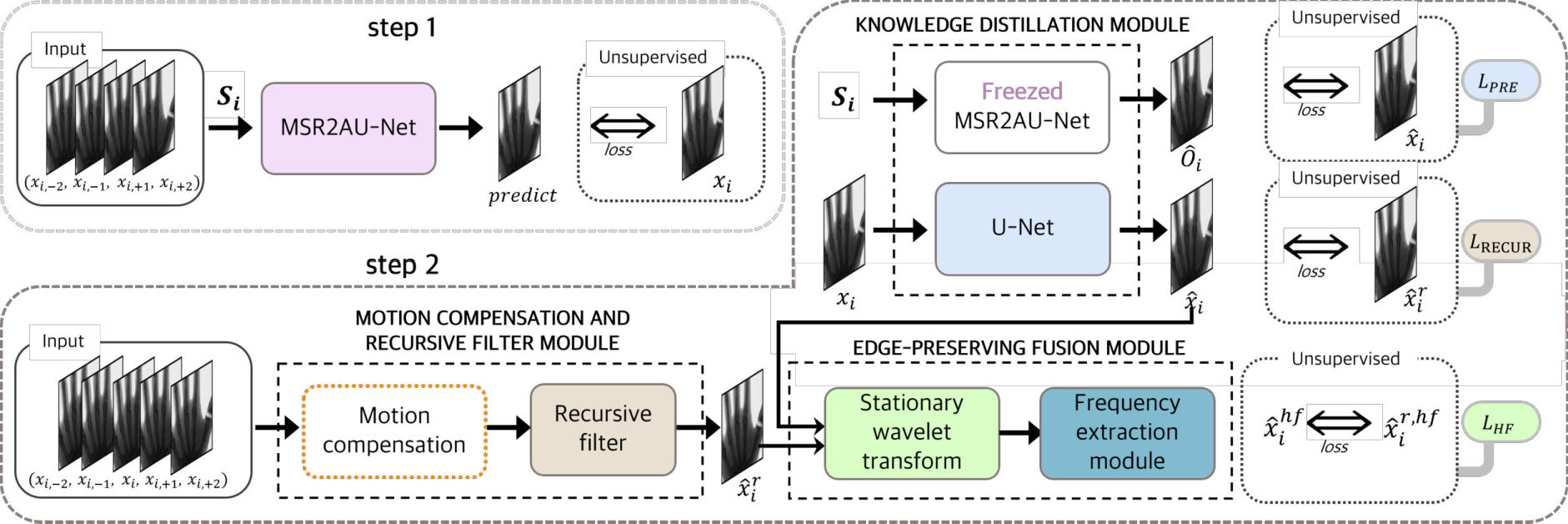}}
   \caption{Schematic diagram of the proposed two-step training framework.}
\label{Fig2} 
\end{figure}

An overview of the proposed approach is provided in Fig.~\ref{Fig2}. We adopted an unsupervised strategy to train the framework without a clean target and introduced two-step training of a dynamic context-aware denoising framework to improve the objective and perceptual quality. The details of the proposed approach are elaborated in the following section.

\subsection{First Training Step: Multi-scale Recurrent Attention U-Net (MSR2AU-Net)}
\begin{figure}[ht]
   \centering
   \includegraphics[width=\linewidth]{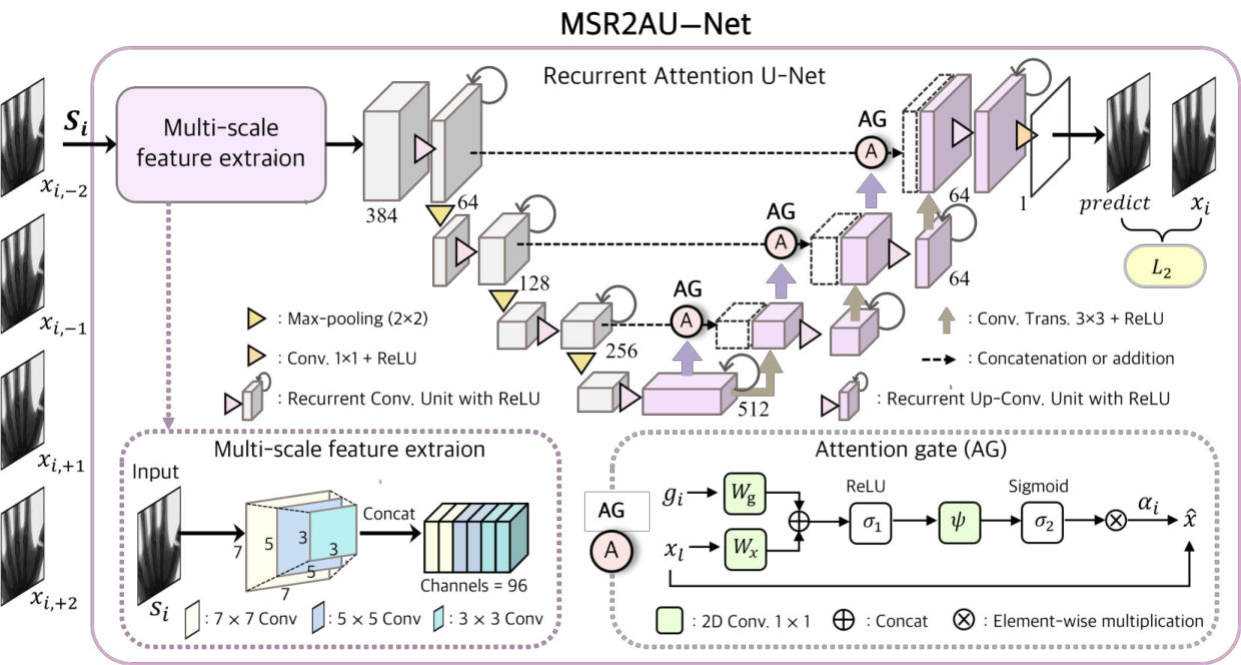}
   \caption{Architectural overview of the multi-scale recurrent attention U-Net (MSR2AU-Net). This figure illustrates the MSR2AU-Net framework employed in the first training step. The network is designed to predict the central frame from noisy X-ray sequences by leveraging multi-scale feature extraction, recurrent layers, and attention mechanisms to enhance denoising performance.\\
   \label{Fig3} 
    } 
\end{figure}

As shown in Fig.~\ref{Fig3}, we train the network to predict the center frame from a multi-frame input to remove the initial noise. We defined a series of consecutive noisy input sets as $S_{i}=[{x_{i,-2},x_{i,-1},x_{i,+1},x_{i,+2}}]$ and $x_{i}$ as the target. This set comprises the $i$-th central frame ($x_{i}$) to be denoised, along with its two preceding frames ($x_{i,-2}$, $x_{i,-1}$) and two following frames ($x_{i,+1}$, $x_{i,+2}$), exhibiting a high degree of anatomical similarity. The use of multiple frames ensures temporal consistency and enhances the model’s ability to reduce noise while preserving essential anatomical details.
Although this might introduce a slight delay, given the high frame rate of typical fluoroscopy systems (15 frames/s) and the slower pace of interventional procedures, the resulting delay is minimal and likely imperceptible to humans, estimated at approximately 133ms. Building on this approach, the Multi-scale Recurrent Attention U-Net (MSR2AU-Net) processes these multi-frame inputs to effectively predict the central frame. Following previous work~\cite{jeon2022mm} highlighting the efficacy of the attention U-Net in noise reduction for LDCT images, we employed the attention U-Net in the MSR2AU-Net architecture. In this study, we replaced standard convolutional units with recurrent residual convolutional units and integrated the attention gate into MSR2AU-Net to enhance denoising performance. In the Recurrent residual convolutional unit~\cite{chung2014empirical}, both recurrent and residual connections are incorporated into each convolutional layer, allowing the network to enhance its feature extraction capabilities without increasing the number of parameters. These units are strategically placed within the encoding and decoding paths of the network to optimize the flow of information.
Additionally, the residual connections help develop deeper and more efficient models by facilitating the flow of gradients during training, thus mitigating the vanishing gradient problem. Thus, incorporating recurrent residual convolutional units allows for iterative information integration and attention-weight computation. This modification improves performance by integrating context information throughout the entire network without introducing additional network parameters\cite{zuo2021r2au,li2022rt}. 

As depicted in Fig.~\ref{Fig3}, MSR2AU-Net comprises multi-scale feature extraction, attention gates, and a recurrent U-Net. During the training phase, we employed convolutional kernels of varied sizes\cite{wang2019multi} with 32 filters for each frame. Each frame is first processed through 3×3 kernels to focus on fine details and local textures, 5×5 kernels to capture mid-range patterns, and 7×7 kernels to extract broader, more global structures within the image. Afterward, the resulting feature maps from these different scales are concatenated, enabling the network to effectively integrate information across multiple levels of detail, ultimately resulting in 384 concatenated feature maps. This methodology allows the network to recognize spatial and temporal correlations across sequential frames, enhancing its ability to extract essential features. We concatenated the feature map channels from each frame to enhance the network capability further. The multi-scale concatenated feature maps are then fed into the Recurrent Attention U-Net, which is trained to predict the center frame. The training process utilizes the $\mathcal{L}_{2}$-norm as the loss function, defined as follows:

\begin{equation}\label{eqn:l2}
\begin{aligned}
\mathcal{L}_{2}(\theta)=
 \left (  x_{i}-F_\theta (S_i)  \right )^2,
\end{aligned}
\end{equation}
where $F_{\theta}$ represents the MSR2AU-Net with parameter $\theta$, and $x_{i}$ is the target.

\subsection{Second Training Step: Edge-preserving Temporal Noise Suppression}
\begin{figure*}[ht]
   \centering
   \includegraphics[width=\linewidth]{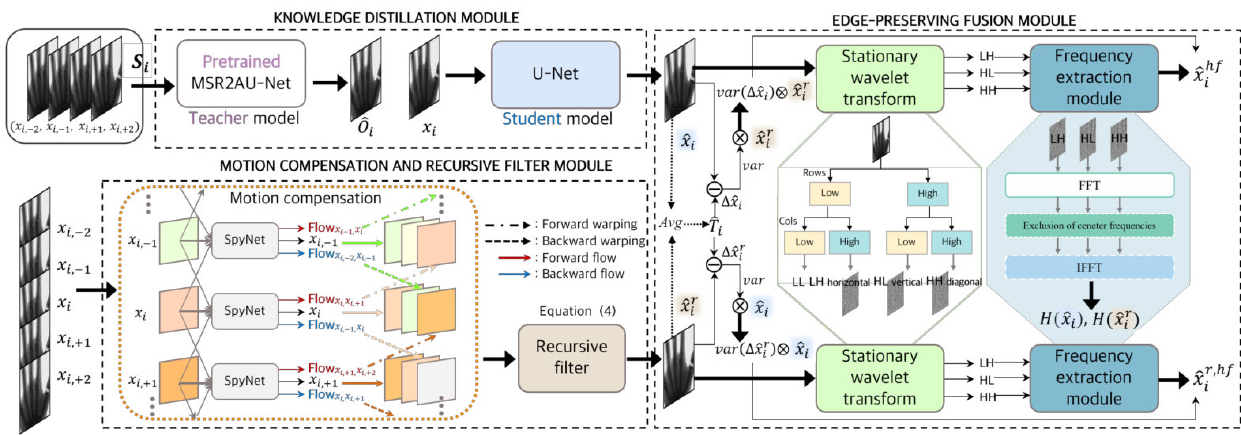}
   \caption{Schematic of the second training step with the network architecture used in the second training phase, where ${\hat{O}_{i}}$ signifies the frozen pre-trained MSR2AU-Net. The outputs $\hat{x}_{i}$ and $\hat{x}_{i}^r$ represent the products of the U-Net and recursive filter, respectively. \textit{Avg} represents the average, and $\hat{T}_{i}$ is the result of the element-wise multiplication of $\hat{x}_{i}$ and $\hat{x}_{i}^r$. The differences $\Delta\hat{x}_i$ and $\Delta\hat{x}_i^r$ are obtained by subtracting $x_i$ and $\hat{x}_i^r$ from $\hat{T}_i$, respectively. ${var}$ is the variance operator. Finally, $\hat{x}_i^{hf}$ is the element-wise multiplication of the high-frequency components of $\hat{x}_i$, denoted as $H(\hat{x}_i)$, with $\text{var}(\Delta\hat{x}_i)\bigotimes \hat{x}_{i}^r$, and $\hat{x}_i^{r,hf}$ is the element-wise multiplication of $H(\hat{x}_i^r)$ with $var(\Delta\hat{x}_i^r)\bigotimes \hat{x}_{i}$.
   \label{Fig4} 
     }
\end{figure*}
The second training step serves two critical objectives: further noise reduction and improved perceptual quality. It ensures the preservation of textural tissue details, even in motion.
\subsubsection{Knowledge Distillation-Based Uncorrelated Noise Suppression Module}

As illustrated in Fig.~\ref{Fig4}, the proposed module employs off-line knowledge distillation, encompassing two distinct models: a teacher model with a pre-trained MSR2AU-Net (in a frozen state) from the first training step and a U-Net-based~\cite{ronneberger2015unet} student model. In the second training step, we used the current frame $x_{i}$ as the input of the U-Net. The teacher model knowledge is transferred to the student model by analyzing the differences in their outputs. The MSE loss measures the distance between the output of the teacher model and the output of the student model. The MSE-based loss function $\mathcal{L}_{pre}$, denoted as the PRE loss, is computed as follows:
\begin{equation}\label{eqn:pre}
\begin{aligned}
\mathcal{L}_{pre} = \sum_{i=1}^{N}\left (\hat{x}_{i}-\hat{O}_{i} \right )^2,
\end{aligned}
\end{equation}
where $\hat{x}_{i}$ and $\hat{O}_{i}$ are the outputs of the student and teacher models, respectively. 
The primary purpose of using the PRE loss $\mathcal{L}_{pre}$ is to eliminate noise from noisy X-ray sequence images effectively. However, this method might not be adequate to attain the desired level of noise reduction, particularly when addressing motion-induced correlated noise.

\subsubsection{Motion Compensation and Recursive Filtering-Based Correlated Noise Suppression Module}

We incorporated a motion compensation approach to bolster the efficacy of reducing motion-induced correlated or structured noise. We can separate a noisy image ${x}_{i}$ it into two components: ${x}_{i}={s}_{i}+{n}_{i}$, where ${s}_{i}$ and ${n}_{i}$ represent the clean signal and its corresponding noise, respectively. Using noisy images, we assembled a collection of similar images from adjacent frames, represented by ${x}_{i}={s}_{i}+{n}_{i}$ and $\hat{x}_{i}={s}_{i}+\delta_{i}+\hat{n}_{i}$, where $\delta_{i}$ symbolizes the difference in the clean signal components in similar images (e.g., slight variations in anatomy across adjacent frames) and $n_{i}$ and $\hat{n}_{i}$ indicate two distinct noise instances. We defined $\theta_{s}$ as the network parameter vector optimized with similar data pairs. We determined $\theta_{s}$ by minimizing the ensuing loss function:
\begin{equation}
\begin{aligned}
    \theta_{s} ={ \small arg \min\limits_{\theta}} \frac{1}{N_{s}}\sum_{i=1}^{N_{s}}||f(s_{i}+n_{i};\theta)-(s_{i}+\delta_{i}+\hat{n}_{i} )||_2^2,
\label{eq:delta-correlated-noise}
\end{aligned}
\end{equation}
where $N_{s}$ represents the count of similar noisy image pairs. With the zero-mean conditional noise ($\mathbb{E}[\hat{n}_{i}|s_{i}+n_{i}]=0$) and the zero-mean conditional discrepancy ($\mathbb{E}[\delta_{i}|s_{i}+n_{i}]=0$), $\theta_{s}$ coincides with network parameters optimized under paired noise-clean images as $N_{s}$ approaches infinity~\cite{niu2022noise}. Consequently, when suitable motion compensation is applied to images from adjacent frames, the value of $\delta_{i}$ decreases substantially, enabling supervised denoising to be conducted using only noisy images. 

As illustrated in Fig.~\ref{Fig4}, we performed motion compensation between adjacent frames within the frame set from $x_{i,-2}$ to $x_{i,+2}$, involving the computation of optical flows between the current frame $x_{i}$ and its neighboring frames ($x_{i,-1}$ and $x_{i,+1}$). The neighboring frames ($x_{i,-1}$ and $x_{i,+1}$) were incorporated into the current frame $x_{i}$ through forward and backward warping. The process is the same for the remaining frames. For flow estimation, we used SpyNet~\cite{ranjan2017optical} pre-trained with ImageNet~\cite{5206848} to estimate the motion.

Subsequently, the motion-compensated output was input into the recursive filtering process, a real-time, edge-preserving smoothing filter~\cite{xu2014image} that effectively removes noise. The recursive filter output can be expressed as follows~\cite{pham2018asymmetric}:
\begin{equation}\label{eqn:recursive}
\begin{aligned}
\hat{x}_{i,j+1}^r = (1-w)\cdot \hat{x}_{i,j}^r + w\cdot{x}_{i,j},
\end{aligned}
\end{equation}
 where $j$ represents the index for ${x}_{i,j}$, taking on values within the range of -2 to 2. The parameter $w$ is a weighting factor between 0 and 1 that determines the influence of the previous output value on the current output value. In this implementation, we set $w=0.2$. Additionally, $\hat{x}_{i,j}^r$ signifies the $j$-th output of the recursive filter. To enhance the signal-to-noise ratio (SNR) and detail preservation, we employed the MSE between the final denoiser's output and the recursive filter's output as the loss function, termed the RECUR loss $(\mathcal{L}_{recur})$. $\mathcal{L}_{recur}$ is defined as follows:
\begin{equation}\label{eqn:recur}
\begin{aligned}
\mathcal{L}_{recur} = \sum_{i=1}^{N}\left ( \hat{x}_{i} - \hat{x}_{i}^r\right )^2,
\end{aligned}
\end{equation}
where $\hat{x}_{i}$ and $\hat{x}_{i}^r$ represent the output of the U-Net and the final recursive filter, respectively. By integrating the PRE loss $\mathcal{L}_{pre}$ with the RECUR loss $\mathcal{L}_{recur}$, we can significantly reduce noise and extract more temporal features between consecutive frames, thereby improving overall denoising performance.

\subsubsection{Edge-preserving Fusion of Uncorrelated and Correlated Noise Reduction Modules}
The edge-preserving fusion module enhances image quality and restores intricate details by suppressing both uncorrelated and correlated noise. The U-Net within the knowledge distillation-based uncorrelated noise suppression module demonstrates robust denoising capabilities. However, it neglects temporal inconsistency over multiple frames due to motion, blurring fine structures. In contrast, the motion compensation and recursive filtering-based correlated noise suppression module may have slightly inferior denoising performance. However, the module guarantees temporal consistency, effectively preserving fine structures. To optimize the strengths of both modules, we design a cross-fusion matrix that generates an ideal output by merging the superior noise removal of the U-Net output with the temporal consistency of the recursive filter output. The process of the cross-fusion matrix is as follows: first, an output called the optimal output is calculated as the average of the U-Net and the recursive filter outputs, as shown below:

\begin{equation}
\hat{T}_i = \frac{1}{2}(\hat{x}_i + \hat{x}_i^r),
\end{equation}
where $\hat{T}_i$ represents the optimal image, $\hat{x}_i$ denotes the U-Net output, and $\hat{x}_i^r$ is the output from the recursive filter.

To maintain intricate texture details, as shown in Fig.~\ref{Fig4}, we can derive two difference images, $\Delta\hat{x}_{i}$ and $\Delta\hat{x}_{i}^r$, from $\hat{T}_i$ to detect changes between the two images, $\hat{x}_{i}$ and $\hat{x}_{i}^r$. This can be formalized as follows:

\begin{equation}
\begin{aligned}
 \Delta\hat{x}_{i} = \left| \hat{T}_{i} - \hat{x}_{i} \right| ,  
\Delta\hat{x}_{i}^r = \left| \hat{T}_{i} - \hat{x}_{i}^r \right|,
\end{aligned}
\end{equation}
where $\Delta\hat{x}_{i}$ represents the level of noise, while $\Delta\hat{x}_{i}^r$ indicates the degree of motion correction. 
 To further refine our analysis, we apply a variance operator to both $\Delta\hat{x}_{i}$ and $\Delta\hat{x}_{i}^r$. The variance operator helps us quantify the variability in these images, allowing us to assess the stability and consistency of the noise reduction and motion correction processes. After obtaining the variance images, we multiply the variance image of $\Delta\hat{x}_{i}^r$ by $\hat{x}_{i}$ to incorporate the noise characteristics into the motion-corrected image. Similarly, we multiply the variance image of $\Delta\hat{x}_{i}$ by $\hat{x}_{i}^r$ to integrate the motion correction details into the noise-reduced image. The fusion matrix outputs $M_{\Delta\hat{x}_{i},\hat{x}_{i}^r}$ and $M_{\Delta\hat{x}_{i}^r,\hat{x}_{i}}$ are defined according to the following equation:
\begin{equation}
\begin{aligned}
M_{\Delta\hat{x}_{i},\hat{x}_{i}^r} = 
    var(\Delta\hat{x}_{i})\otimes\hat{x}_i^r,  
 M_{\Delta\hat{x}_{i}^r,\hat{x}_{i}} = var(\Delta\hat{x}_{i}^r)\otimes\hat{x}_{i},
\end{aligned}
\end{equation}
where ${var}$ denotes the variance operator and $\otimes$ represents element-wise multiplication. This approach allows us to dynamically balance the strengths of both modules, resulting in a more robust and optimized final output.

\subsubsection{High-Frequency Extraction and Enhancement using Wavelet and Fourier Transform Module}
To retain intricate texture details in fluoroscopic images affected by motion blur, we additionally incorporated a stationary wavelet transform (SWT) module to decompose the two outputs ($\hat{x}_i$ and $\hat{x}_{i}^r$) into four subbands each. We performed the Level 1 SWT using the Haar function as the wavelet function. One subband contains low-frequency information, whereas the other three capture high-frequency information. However, while SWT effectively preserves multi-resolution details, it may not fully capture the critical high-frequency components necessary for fine texture restoration. 

To address this limitation, we further enhanced our approach by incorporating a Frequency Extraction Module based on Fast Fourier Transform (FFT). As shown in Fig~\ref{Fig4}, The three high-frequency subbands, excluding the low-frequency subband, from both outputs ($\hat{x}_i$ and $\hat{x}_{i}^r$) are then passed through the Frequency Extraction Module. In this module, each of the three high-frequency subbands undergoes FFT to extract high-frequency components further. Subsequently, the low-frequency components near the center of the spectrum are excluded, and an inverse FFT (iFFT) is performed to reconstruct the high-frequency details. The three transformed subbands are then combined to form the images, denoted as H($\hat{x}_i$) and H($\hat{x}_{i}^r$), with the extracted high-frequency characteristics. H($\hat{x}_i$) and H($\hat{x}_{i}^r$) are then concatenated with the fusion matrix outputs $M_{\Delta\hat{x}_{i},\hat{x}_{i}^r}$ and $M_{\Delta\hat{x}_{i}^r,\hat{x}_{i}}$, respectively, and multiplied. The high-frequency components $\hat{x}_{i}^{hf}$ and $\hat{x}_{i}^{r,hf}$ are then formalized as follows:
\begin{equation}
\hat{x}_{i}^{hf} = M_{\Delta\hat{x}_{i},\hat{x}_{i}^r} \otimes H(\hat{x}_i),
 \hat{x}_{i}^{r,hf} = M_{\Delta\hat{x}_{i}^r,\hat{x}_{i}} \otimes H(\hat{x}_{i}^r),
\end{equation} 
where $\otimes$ denotes element-wise multiplication. Subsequently, we determined the high-frequency-based perceptual loss. The high-frequency loss function, referred to as the HF loss ($\mathcal{L}_{hf}$), sums the squared differences between the high-frequency components of the two outputs, $\hat{x}_{i}^{hf}$ and $\hat{x}_{i}^{r,hf}$, and is defined as follows:
\begin{equation}\label{eqn:high}
\begin{aligned}
\mathcal{L}_{hf} = \sum_{i=1}^{N}\left ( \hat{x}_{i}^{hf}- \hat{x}_{i}^{r,hf}\right )^2.
\end{aligned}
\end{equation}

\subsubsection{Total Loss Function}
Finally, we defined the total loss function, denoted as $\mathcal{L}_{final}$, for training the U-Net-based final denoising network. It combines the PRE loss $\mathcal{L}_{pre}$, the RECUR loss $\mathcal{L}_{recur}$, and the high-frequency loss $\mathcal{L}_{hf}$ as follows:
\begin{equation}\label{eqn:final}
\begin{aligned}
\mathcal{L}_{final} = \mathcal{L}_{pre}+ \alpha\cdot\mathcal{L}_{recur}+\mathcal{L}_{hf} .
\end{aligned}
\end{equation}
The hyperparameter $\alpha$ balances the contributions of the uncorrelated and correlated noise suppression modules. During training, we set $\alpha$ to 1.

\section{EXPERIMENTS}
\subsection{Experimental Setup}
\subsubsection{Datasets}
This study utilizes several datasets, including two dynamic phantom datasets and a clinical in vivo dataset. We procured fluoroscopy images of two dynamic phantoms at 60 kVp for model training and evaluation, with low- and high-dose settings of 0.002 and 0.02 mAs, respectively. One phantom replicated a real-bone X-ray hand; the other represented a needle biopsy with spherical objects. Their motions were simulated using the Model 008A Dynamic Thorax Phantom (CIRS, Norfolk, VA, USA) with translations up to $\pm 2$ cm. The dynamic phantom dataset comprises a total of 3,500 images with various motions. In our experiment, 2,400 pairs of low- and high-dose images were used for training, while the remaining 1,100 images were used for testing. From the training images, we randomly extracted 64$\times$64 patches from 700$\times$700 images. A 5$\%$ subset of this training dataset was randomly chosen for validation. Furthermore, we evaluated the proposed method's performance on 350 clinical images obtained from spinal surgery patients, with C-arm settings at 106 kV and 0.02 mAs. For these clinical images, we employed cross-validation to demonstrate
the model’s robustness against overfitting and its generalization to unseen data. Additionally, we utilized the publicly available ``2016 NIH-AAPM-Mayo Clinic Low Dose CT Grand Challenge'' dataset~\cite{mayochallenge}, which includes NDCT and LDCT images from ten patients. Each pair consists of 1-mm thickness 512$\times$512 images. For this dataset, we used data from eight patients for training (4,800 images) and the other two for testing (1,136 images).

\subsubsection{Implementation Details}
The proposed network and benchmark algorithms were implemented using PyTorch, ensuring consistency by configuring all networks with the same settings. We used the Adam optimizer~\cite{kingma2014adam} with a learning rate of 1$\times10^{-4}$. Additionally, we limited the maximum number of training epochs to 200. The first and second training steps were performed by randomly extracting patches of size 64$\times$64 pixels from the input data.

\subsubsection{Compared Methods}
19 state-of-the-art methods were used for comparison: seven unsupervised models (BM3D~\cite{dabov2007image}, NLM~\cite{li2014adaptive}, Noisier2Noise (N2N)~\cite{moran2020noisier2noise}, Noise2Void (N2V)~\cite{krull2019noise2void}, Noisy-As-Clean (N2C)~\cite{xu2020noisy}, MM-Net~\cite{9963593}, and DenoisingGAN~\cite{kim2024unsupervised}) and 12 supervised models (U-Net~\cite{ronneberger2015unet}, RED-CNN~\cite{7947200}, WGAN-VGG~\cite{8340157}, DnCNN~\cite{zhang2017beyond}, MCDnCNN~\cite{wu2020combined}, UDDN~\cite{luo2020ultra}, EEDN~\cite{luo2022edge}, FastDVDNet~\cite{van2021real}, CCN-CL~\cite{tang2022ccn}, StruNet~\cite{ma2023strunet}, CTformer~\cite{wang2023ctformer}, and Kim~\textit{et al.}~\cite{kim2022wavelet}). To ensure a fair comparison, all experiments were conducted under identical conditions, with both quantitative and qualitative results reflecting the direct output of the models without any preprocessing. We visually analyzed the qualitative results by examining different denoising predictions from each model.

\subsubsection{Quantitative Evaluation Metrics}
Five metrics were used to quantitatively evaluate the algorithm performance, including peak signal-to-noise ratio (PSNR), structural similarity index measure (SSIM), efficient deep-detector image quality assessment (EDIQA)~\cite{lee2023efficient}, natural image quality evaluator (NIQE)~\cite{mittal2012making}, and visual information fidelity (VIF)~\cite{sheikh2006image}.
\subsection{Results on Dynamic Phantom Datasets}
\begin{figure*}[ht]
\centering
\includegraphics[width=\linewidth]{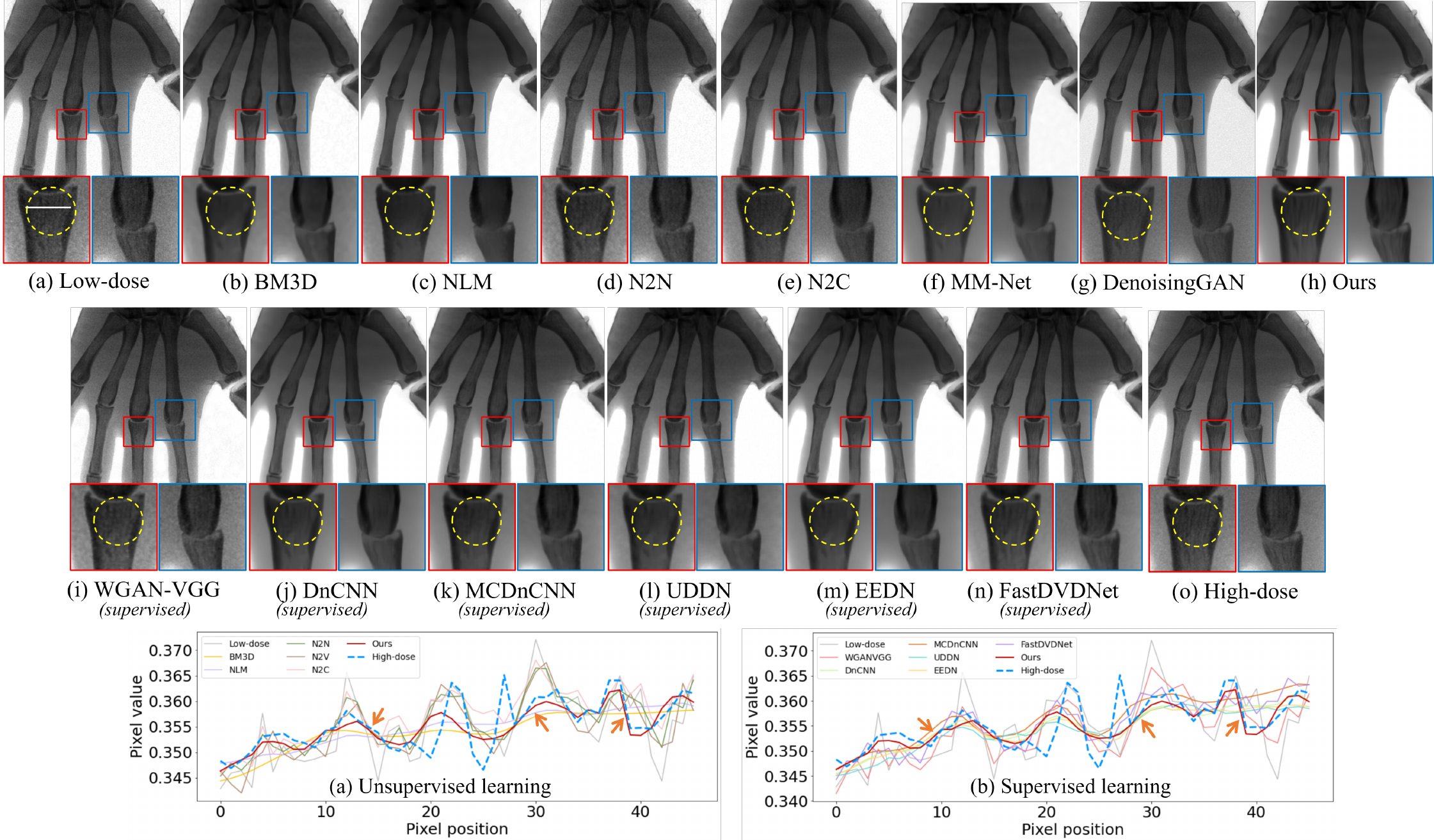}
\caption{Comparative denoising results on the dynamic anthropomorphic hand phantom dataset with various networks. The line profiles in the third row are plotted along the white line within the red region of interest.}
\label{Fig5}
\end{figure*}

\begin{figure*}[ht]
\centering
\includegraphics[width=\linewidth]{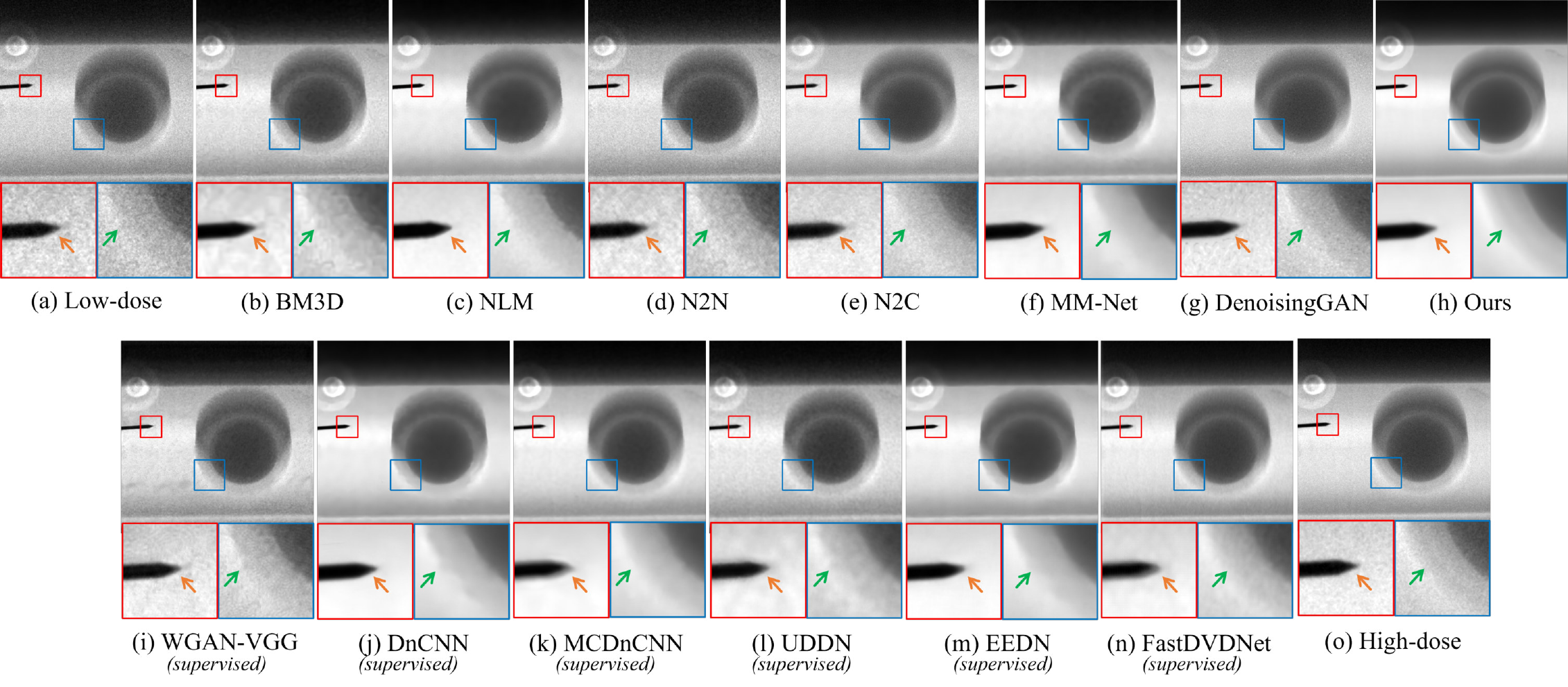}
\caption{Denoised outcomes from various networks applied to the dynamic needle tip with a spherical lesion phantom dataset.}
\label{Fig6}
\end{figure*}

\begin{table}[ht]
\renewcommand{\arraystretch}{1.2}
\centering
\caption{Comparative performance metrics of various methods across two dynamic phantom datasets. The presented figures represent aggregated results from both datasets. The best results among unsupervised methods are marked in bold.}
\resizebox{\linewidth}{!}{%
\begin{tabular}{cc|c|c|c|c|c}
\toprule[1pt]
\multicolumn{2}{c|}{Method}                                       & SSIM $\uparrow$ & PSNR [dB]$\uparrow$   & EDIQA$\uparrow$   & NIQE $\downarrow$    & VIF $\uparrow$      \\ \hline
\multicolumn{1}{c|}{\multirow{9}{*}{\rotatebox{90}{Unsupervised}}} & Low-dose     &0.9353 $\pm$ 0.0046  & 36.08 $\pm$ 4.6005 & 0.6334 &7.6476 &0.1803      \\
\multicolumn{1}{c|}{}                              & BM3D         & {0.9769 $\pm$ 0.0041}  & {38.25 $\pm$ 6.4440}   &0.6440  &7.0981 & 0.2454    \\
\multicolumn{1}{c|}{}                              & NLM        & 0.9767 $\pm$ 0.0028           & 37.98 $\pm$ 6.1315   &0.6359  &7.7627 &0.2311 \\
\multicolumn{1}{c|}{}                              & N2N        & 0.9633 $\pm$ 0.0040           & 37.10 $\pm$ 5.4677  &0.6306  &7.2206 &0.2115    \\ 
\multicolumn{1}{c|}{}                              & N2V        & 0.9576 $\pm$ 0.0046           & 37.07 $\pm$ 5.2784   &0.6334   &7.0611 &0.2157  \\
\multicolumn{1}{c|}{}                              & N2C        & 0.9664 $\pm$ 0.0033           & 37.84 $\pm$ 5.4095    &0.6357  &7.0035 &0.2313  \\
\multicolumn{1}{c|}{}                              & MM-Net        & 0.9779 $\pm$ 0.0021           & 38.61 $\pm$ 5.2552   &0.6437   &6.0476 &0.2552  \\
\multicolumn{1}{c|}{}                              & DenoisingGAN    & 0.9770 $\pm$ 0.0030           & 38.27 $\pm$ 5.5501   &0.6420 &6.1758 &0.2359   \\
\multicolumn{1}{c|}{}                              & Ours       & \textbf{0.9803 $\pm$ 0.0023}  & \textbf{39.12 $\pm$ 5.2741}      & \textbf{0.6447} &\textbf{5.4028} &\textbf{0.2696}\\ \hline
\multicolumn{1}{c|}{\multirow{12}{*}{\rotatebox{90}{Supervised}}}   & U-Net      & 0.9788 $\pm$ 0.0025           & 44.27 $\pm$ 2.1220      &  0.6409 &5.5063 &0.2839 \\
\multicolumn{1}{c|}{}                              & RED-CNN    & 0.9779 $\pm$ 0.0025           & 43.50 $\pm$ 1.8464   &0.6427    &5.5039 &0.2659 \\
\multicolumn{1}{c|}{}                              & WGAN-VGG    & 0.9607 $\pm$ 0.0045           & 38.89 $\pm$ 0.8627   &  0.6436  &6.1488 &0.2650   \\ 
\multicolumn{1}{c|}{}                              & DnCNN      & 0.9769 $\pm$ 0.0024           & 43.56 $\pm$ 1.9354  & 0.6465   &5.4705 &0.2758  \\
\multicolumn{1}{c|}{}                              & MCDnCNN    & 0.9745 $\pm$ 0.0064           & 43.59 $\pm$ 1.6462   & 0.6451  &6.0565 &0.2591  \\
\multicolumn{1}{c|}{}                              & UDDN       & 0.9767 $\pm$ 0.0038           & 43.53 $\pm$ 2.0156  & 0.6430    &6.1024 &0.2548  \\ 
\multicolumn{1}{c|}{}                              & EEDN       & 0.9775 $\pm$ 0.0019           & 43.79 $\pm$ 1.9101  & 0.6450   &6.0479 &0.2644  \\ 
\multicolumn{1}{c|}{}                              & FastDVDNet & 0.9768 $\pm$ 0.0079           & 42.57 $\pm$ 2.1292  & 0.6378   &5.9552 &0.2593  \\ 
\multicolumn{1}{c|}{}                              & CCN-CL    & 0.9801 $\pm$ 0.0034            & 43.97 $\pm$  2.1929  &0.6495 &5.5933 &0.2691   \\
\multicolumn{1}{c|}{}                              & CTformer       & 0.9771 $\pm$ 0.0029            & 42.98 $\pm$  1.6643   &0.6448 &5.4383 &0.2634  \\ 
\multicolumn{1}{c|}{}                              & Kim~\textit{et al.} 2022 & 0.9794 $\pm$ 0.0021            & 44.17 $\pm$  2.0456   &0.6450  &5.3020 &0.2738  \\ 

\bottomrule[1pt]
\end{tabular}%}
\label{Table1}
}\vspace{-0.18cm}
\end{table}

We conducted experiments on the dynamic phantom dataset, with visual results for the anthropomorphic hand phantom in Fig.~\ref{Fig5}. The top and bottom rows display the denoising outcomes of representative unsupervised and supervised methods, respectively. All denoising networks effectively generated denoised images. Among the unsupervised approaches, BM3D and NLM  seem to suppress the noise, but due to the blurry appearance, they struggle to preserve the fine bone tissue structures within the red region of interest (ROI) on the left, as indicated by the circular dashed yellow lines. Line profiles highlighted the challenge of detail preservation along the white lines in the left ROI, appearing excessively smooth compared to actual high-dose line profiles. N2N and N2V inadequately removed noise, while N2C achieved some noise removal, resulting in the smearing of bone structure boundaries within the blue ROI. Although MM-Net demonstrated strong performance in artifact suppression, as seen in the red and blue ROIs, it produced overly smoothed denoised images. This over-smoothing occurs because the network's performance, driven by its multipatch and multimask matching loss, prioritizes overall image quality, effectively preserving larger structures but at the cost of losing fine textures and details. On the other hand, DenoisingGAN achieved better perceptual quality due to the combination of GAN and perceptual losses. Furthermore, the inclusion of wavelet loss in the loss function enhances its perceptual quality, making it superior to MM-Net in this aspect. It can be observed that DenoisingGAN better mitigates the over-smoothing problem compared to MM-Net. In contrast to the residual artifacts seen in DenoisingGAN denoised images, the proposed unsupervised method effectively reduced noise and enhanced fine structure visibility. Examining the denoising results of the supervised methods in the second row reveals that MCDnCNN, EEDN, and FastDVDNet excelled in noise reduction. However, MCDnCNN and EEDN tended to produce overly smooth images, obscuring fine details. The proposed method showed better outcomes than FastDVDNet, excelling in noise reduction and the restoration of anatomical details within the zoomed-in ROIs, with profiles closely matching those of high-dose images, as highlighted by orange arrows.

Fig.~\ref{Fig6} exhibits the denoising outcomes of prominent unsupervised and supervised methods applied to a dynamic phantom simulating a needle biopsy with spherical lesions. Mirroring the results observed in the hand phantom, the proposed method continues to demonstrate the best denoising performance among unsupervised techniques. MM-Net failed to preserve sharp edges, as seen by the green arrow in the blue ROI, and DenoisingGAN struggled with artifact suppression, introducing unwanted white dot artifacts. Compared with supervised learning methods, both MCDnCNN and FastDVDNet exhibited motion artifacts, highlighted by the orange arrow within the red ROI. Conversely, DnCNN struggled to preserve sharp edges, as indicated by the green arrow in the blue ROI. In contrast, the proposed method effectively preserved the edges of moving objects, underscoring its superiority in maintaining crucial details. This is due to the integration of a motion compensation module and an edge-enhancing cross-fusion module, which ensures effective noise reduction and precise motion flow processing. This preservation is evident through the clearly restored contours of the moving needle and spherical lesions within the red and blue ROIs, devoid of motion blurring.

The quantitative outcomes of our statistical analysis, focusing on the dynamic anthropomorphic hand phantom and a needle with a spherical lesion phantom dataset, are summarized in Table~\ref{Table1}. The best results are highlighted in bold. our proposed algorithm demonstrates superior performance among unsupervised learning-based methods in terms of PSNR. Notably, when assessed by the SSIM, which is more closely correlated with human perceptual quality~\cite{sara2019image}, the proposed algorithm outperformed all other evaluated algorithms, including those based on supervised learning, which further emphasizes the efficacy of our approach. The proposed method effectively addresses the over-smooth issue observed with MM-Net and preserves fine structural details. The improvement in PSNR is primarily due to the recursive-based loss and the cross-fusion matrix's effective combination of noise reduction and temporal consistency. Furthermore, as indicated in Table~\ref{Table1}, the EDIQA, NIQE, and VIF values of the proposed method are the highest among the unsupervised methods. VIF measures the similarity between the reference and degraded images using natural scene statistics and provides an accurate surrogate for subjective image quality as assessed by radiologists~\cite{8839547, 9669891, 9726904}. Similarly, EDIQA and NIQE have shown a significant correlation with radiologists’ scores~\cite{lee2022no,lee2023efficient}. Remarkably, despite not using clean images, our method's NIQE value surpassed most state-of-the-art algorithms, with the exception of Kim~\textit{et al.}~\cite{kim2022wavelet}, while the VIF score exceeded those of several supervised methods, including RED-CNN, WGAN-VGG, MCDnCNN, UDDN, EEDN, FastDVDNet, CCN-CL, and CTformer. These perceptual quality metrics, closely correlated with clinical image assessments, demonstrate that our method not only closely resembles NDCT images but also meets high diagnostic standards. In conclusion, the proposed model excels in both noise reduction and structure preservation, aligning with the perceptual quality standards valued by radiologists. This indicates that our enhanced images are not only technically superior but also diagnostically valuable.

\subsection{Cross-dataset Evaluation on the Clinical In Vivo Dataset} 
\begin{figure}[ht!]
\centering
\includegraphics[width=\linewidth]{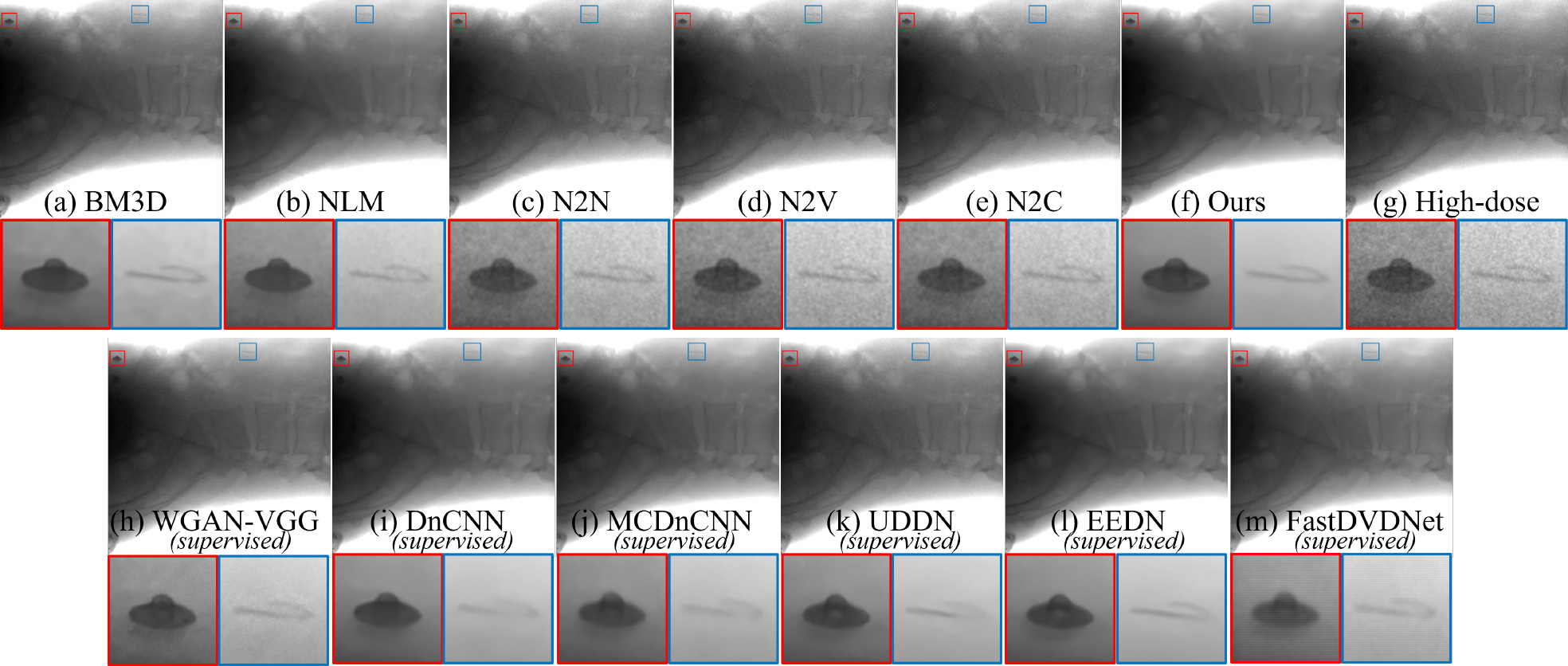}
\caption{Qualitative denoising comparisons on the in vivo dataset using various networks.}
\label{Fig7}
\end{figure}
To further evaluate the performance of the proposed method, cross-dataset evaluations were conducted using distinct datasets for training and testing. All algorithms were trained on the mentioned phantom datasets and evaluated on an in vivo dataset from a patient undergoing spinal surgery. Fig.~\ref{Fig7} depicts the representative results from applying various algorithms to clinical data. Each denoising algorithm demonstrated varying degrees of noise suppression efficacy. However, residual noise levels remained comparatively high in the results of N2N and N2V, as presented in the magnified sections of Fig.~\ref{Fig7}. Conversely, such methods as BM3D and NLM excelled in noise removal compared to their unsupervised counterparts. Nonetheless, as indicated by the blue ROI, BM3D produced distorted background noise textures, whereas NLM resulted in blurry wire edges.

Consistent with the findings from the dynamic phantom images, supervised learning-based approaches outperformed their unsupervised counterparts in denoising efficiency. However, DnCNN, UDDN, and FastDVDNet exhibited excessively blurry edges in the highlighted red and blue boxes, leading to the loss of fine structures (e.g., broken wires). Moreover, MCDnCNN and FastDVDNet introduced significant motion blur, while WGAN-VGG left residual noise. In contrast, the proposed method efficiently reduced image noise and artifacts while excellently preserving structural details. It notably outperformed supervised methods in retaining fine details, underscoring its superior capability in detail preservation.

\subsection{CT MAYO DATASET}
\begin{table}[ht]
\renewcommand{\arraystretch}{1.2}
\centering
\caption{COMPARATIVE PERFORMANCE METRICS OF VARIOUS METHODS USING MAYO CLINIC DATASETS. THE BEST RESULTS AMONG UNSUPERVISED METHODS ARE MARKED IN BOLD.}
\resizebox{\linewidth}{!}{%
\begin{tabular}{cc|c|c|c|c|c}
\toprule[1pt]
\multicolumn{2}{c|}{Method MAYO}                                       & SSIM $\uparrow$ & PSNR [dB]$\uparrow$          & EDIQA$\uparrow$   & NIQE $\downarrow$    & VIF $\uparrow$    \\ \hline
\multicolumn{1}{c|}{\multirow{9}{*}{\rotatebox{90}{Unsupervised}}} & Low-dose     &0.8774 $\pm$ 0.0416  & 31.76 $\pm$ 1.9024            & 0.4244               & 4.9400                & 0.3031\\
\multicolumn{1}{c|}{}                              & BM3D         & {0.8473 $\pm$ 0.0398}  & {35.28 $\pm$ 1.9624}       & 0.6502                & 4.8167                & 0.2905\\
\multicolumn{1}{c|}{}                              & NLM        & 0.8435 $\pm$ 0.0351           & 35.32 $\pm$ 1.8458     & 0.6089                & 4.1086                & 0.2860\\
\multicolumn{1}{c|}{}                              & N2N        & 0.8272 $\pm$ 0.0379           & 33.61 $\pm$ 1.9975     & 0.5251                & 4.8817                & 0.2485\\
\multicolumn{1}{c|}{}                              & N2V        & 0.8281 $\pm$ 0.0287           & 32.72 $\pm$ 1.5709      & 0.4591                & 5.4083                & 0.3153\\
\multicolumn{1}{c|}{}                              & N2C        & 0.8408 $\pm$ 0.0387           & 34.69 $\pm$ 1.7333       & 0.6306                & 4.2492                & 0.3235\\
\multicolumn{1}{c|}{}                              & MM-Net     & 0.9527 $\pm$ 0.0189           & 36.37 $\pm$ 1.6849        & 0.6778                & 3.9261                & 0.3756\\
\multicolumn{1}{c|}{}                              & DenoisingGAN    & 0.9512 $\pm$ 0.0195           & 36.02 $\pm$ 1.7159        & 0.6499                & 3.7440                & 0.3507\\
\multicolumn{1}{c|}{}                              & Ours       & \textbf{0.9591 $\pm$ 0.0161}  & \textbf{36.62 $\pm$ 1.6851}        & \textbf{0.7050}                & \textbf{3.5426}                 &\textbf{0.3807} \\ \hline
\multicolumn{1}{c|}{\multirow{12}{*}{\rotatebox{90}{Supervised}}}   & U-Net      & 0.9612 $\pm$ 0.0143           & 36.74 $\pm$ 1.6004        & 0.7106                & 4.6652                & 0.3779\\
\multicolumn{1}{c|}{}                              & RED-CNN    & 0.9590 $\pm$ 0.0144           & 37.02 $\pm$ 1.7020        & 0.7102                & 4.3603                & 0.3890\\
\multicolumn{1}{c|}{}                              & WGAN-VGG    & 0.9470 $\pm$ 0.0311            & 35.66 $\pm$  1.8239        & 0.6364                & 4.0734                & 0.3600\\ 
\multicolumn{1}{c|}{}                              & DnCNN      & 0.9613 $\pm$ 0.0146            & 36.95 $\pm$  1.7202       & 0.7071                & 4.3133                & 0.3868\\
\multicolumn{1}{c|}{}                              & MCDnCNN    & 0.9552 $\pm$ 0.0284            & 36.59 $\pm$ 1.6862       & 0.7080               & 4.3342                & 0.3789\\
\multicolumn{1}{c|}{}                              & UDDN       & 0.9589 $\pm$ 0.0145           & 36.40 $\pm$ 1.7064          & 0.6992                & 4.7336                & 0.3580\\ 
\multicolumn{1}{c|}{}                              & EEDN       & 0.9603 $\pm$ 0.0142            & 36.58 $\pm$ 1.6956       & 0.7028                & 3.5345                & 0.3760\\ 
\multicolumn{1}{c|}{}                              & FastDVDNet & 0.9584 $\pm$ 0.0154            & 36.57 $\pm$ 1.6848       & 0.6788                & 3.4606                & 0.3744\\ 
\multicolumn{1}{c|}{}                              & CCN-CL   & 0.9610 $\pm$ 0.0150            & 36.91 $\pm$  1.6611      & 0.7023                & 3.3453                & 0.3801\\
\multicolumn{1}{c|}{}                              & STRUNet       & 0.9579 $\pm$ 0.0160            & 36.74 $\pm$  1.6983        & 0.6986                & 3.1454                & 0.3810\\ 
\multicolumn{1}{c|}{}                              & CTformer       & 0.9608 $\pm$ 0.0162            & 36.90 $\pm$  1.6802      & 0.7040                & 3.1930                & 0.3913\\ 
\multicolumn{1}{c|}{}                              & Kim~\textit{et al.} 2022 & 0.9611 $\pm$ 0.0151            & 36.85 $\pm$  1.7048       & 0.7047                & 3.7005                & 0.3878\\ 

\bottomrule[1pt]
\end{tabular}%}
\label{Table1_revision}
}\vspace{-0.18cm}
\end{table}

\begin{figure}[ht!]
\centering
\includegraphics[width=\linewidth]{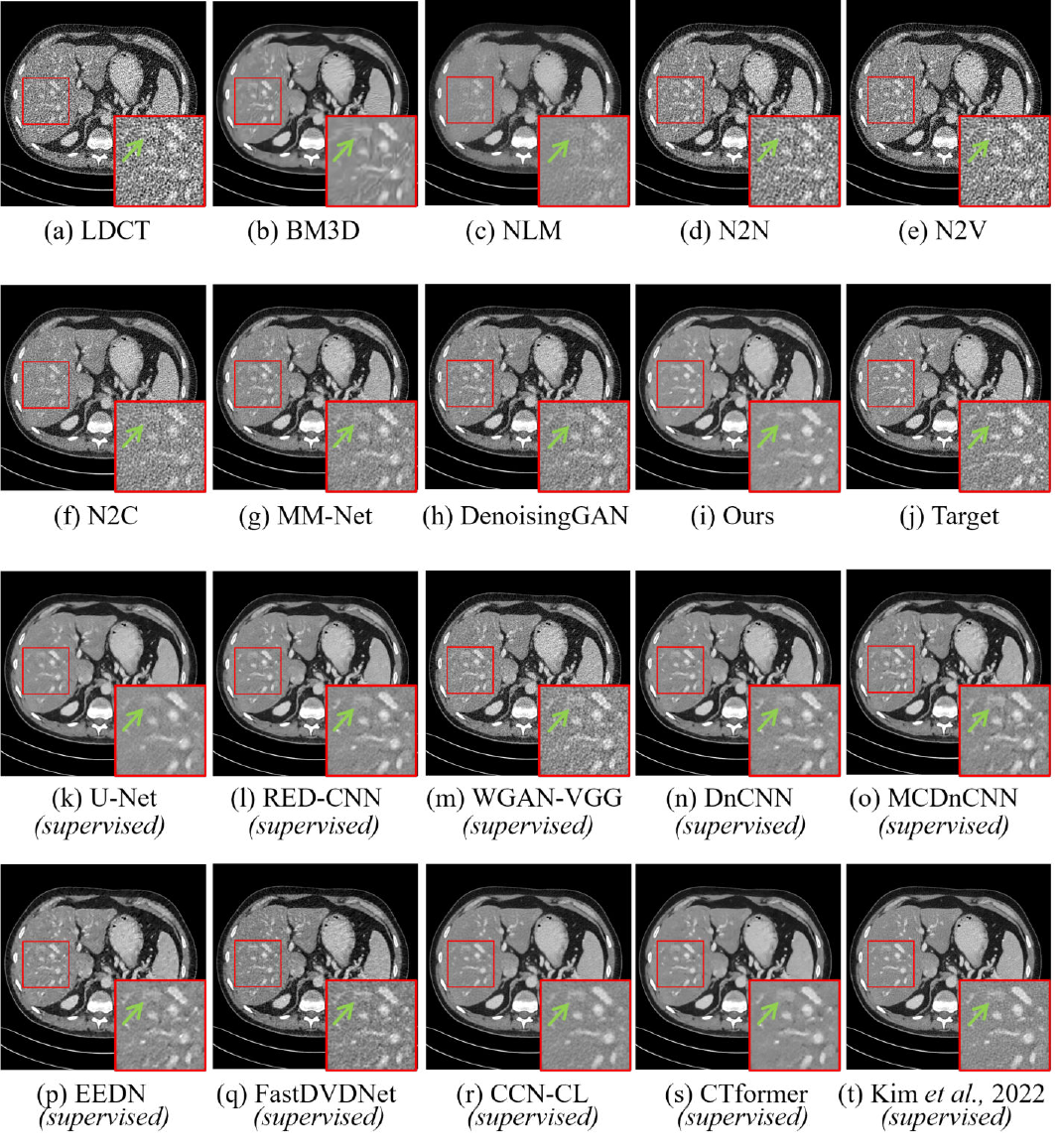}
\caption{Qualitative results of CT images denoised using various methods trained on the Mayo Clinic dataset. The display window is [-150, 250] HU. The red ROI is enlarged for detailed visual comparison.}
\label{Figct1}
\end{figure}

\begin{figure}[ht!]
\centering
\includegraphics[width=\linewidth]{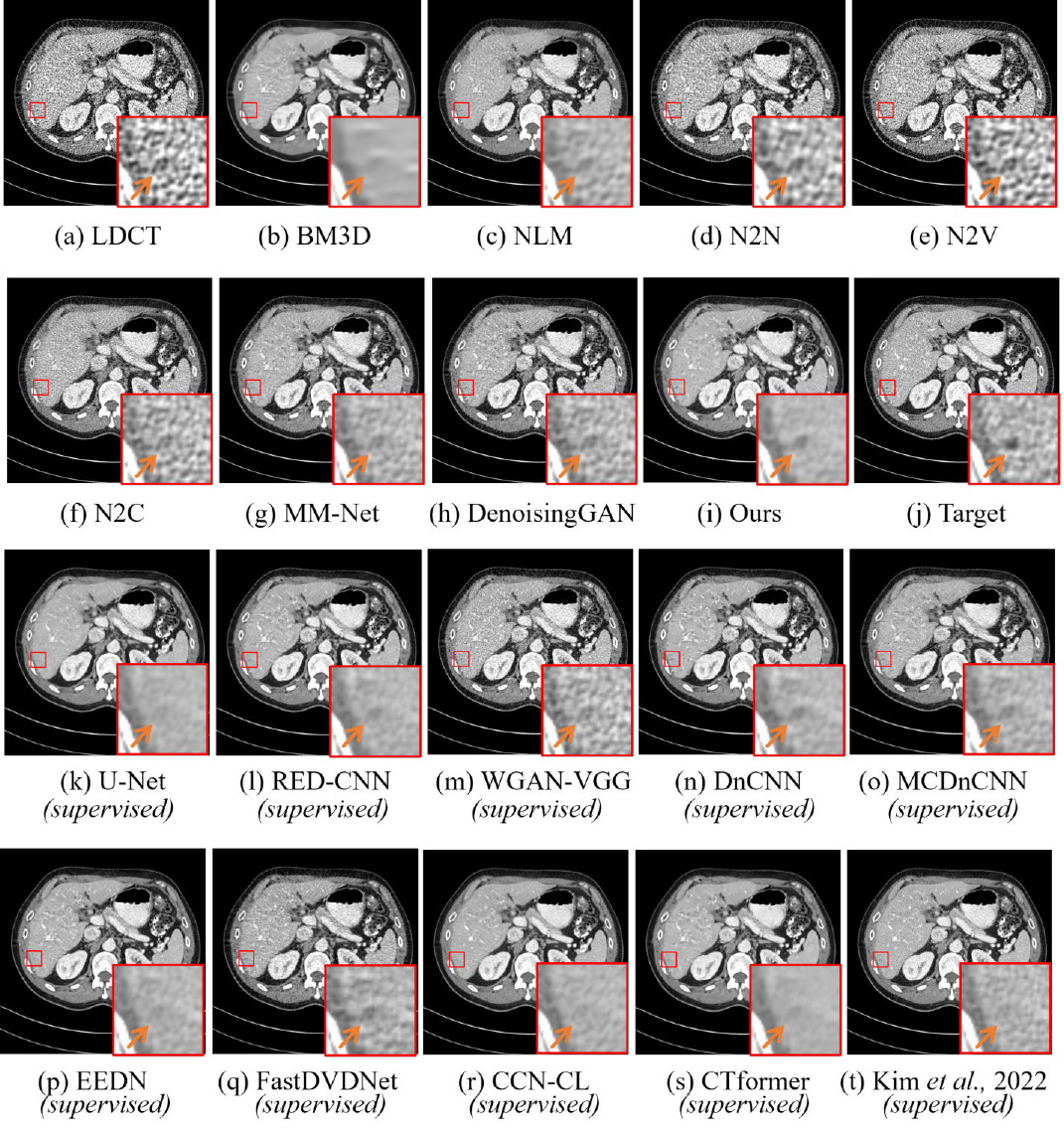}
\caption{Qualitative results of lesion areas in CT images denoised using various methods trained on the Mayo Clinic dataset. The display window is [-150, 250] HU. The red ROI is enlarged for comparison, and the orange arrow indicates a lesion.}
\label{Figct2}
\end{figure}
Fig.~\ref{Figct1} and Fig.~\ref{Figct2} show denoised CT images using various methods, with display windows set to [-150, 250] HU. As seen in Fig.~\ref{Figct1} (a), (j) and Fig.~\ref{Figct2}(a), (j), low-density lesions and tissue structures are harder to identify in LDCT images compared to NDCT images. Denoising methods show varying success in reducing noise and artifacts in LDCT images. BM3D reduces noise but suffers from over-smoothing (Fig. \ref{Figct1}(b)), while NLM loses texture details. MM-Net and DenoisingGAN, which combine a wavelet-based loss with MSE, outperform N2N, N2V, and N2C in preserving structural details and enhancing denoising performance, as indicated by the zoomed red ROI (Fig. \ref{Figct1}(g), (h)). However, some residual noise still remains. Our proposed algorithm shows significantly better performance in artifact suppression and detail preservation. Among supervised methods, U-Net and CCN-CL produce over-smoothed results, leading to blurred edges in fine tissue details. Both WGAN-VGG and FastDVDNet introduce noticeable artifacts into the denoised images. In contrast, RED-CNN, DnCNN, and CT-Former demonstrate superior denoising performance while maintaining structural integrity. When compared to CT-Former, our proposed unsupervised method closely resembles NDCT images in restoring fine structures, as highlighted by the green arrows.

To evaluate lesion preservation, we examined the denoised regions containing lesions, as shown in Fig. \ref{Figct2}. BM3D results in the loss of the lesion's shape, while N2N and N2V leave the noise. In contrast, the other methods produce sharper edges, with MM-Net, DenoisingGAN, and our proposed method standing out, particularly in Fig. \ref{Figct2}(i), where the lesion boundary is clearly defined. Compared to the supervised methods, U-Net, EEDN, CCN-CL, CT-Former, and Kim~\textit{et al}.'s method struggles to make the lesions visible. However, DnCNN, MCDnCNN, and FastDVDNet show slightly better lesion visibility. Our method offers clearer delineation of small lesions, making tiny structures as identifiable as in NDCT images, as shown by the orange arrows.

Table~\ref{Table1_revision} presents the quantitative evaluations of all the denoising methods, with the best metrics highlighted in bold. It is clear that MM-Net, DenoisingGAN, and the proposed method perform better than BM3D through N2C across five different metrics. Among the CNN-based methods, our proposed method achieves the highest scores in all metrics among the unsupervised approaches. Furthermore, the average PSNR/SSIM values of the proposed method surpass those of supervised methods such as WGAN-VGG, MCDnCNN, UDDN, EEDN, and FastDVDNet. In terms of perceptual metrics, our proposed method with wavelet and fourier transform-based edge-preserving loss shows higher NIQE and VIF values compared to MM-Net, which is based on wavelet loss, and DenoisingGAN, which uses perceptual loss. Compared to supervised methods, U-Net and RED-CNN achieved the highest EDIQA scores. For NIQE, the latest algorithms like CTformer and CCN-CL generally achieved higher scores, with our proposed method also performing competitively.

\subsection{Ablation Study of the Proposed Method}
\label{sec:ablation-study}
\begin{table*}
\renewcommand{\arraystretch}{1.2}
\caption{Quantitative ablation study results indicating component contributions to the proposed unsupervised denoising framework. Model configurations are denoted by numbers in parentheses (e.g., (1)), with top performances highlighted in bold.}
\label{Table2}
\centering
\resizebox{\textwidth}{!}{%
\begin{tabular}{c||c|ccc|c|c|c|c|c|c||c|c}
\toprule[1pt]
\multirow{2}{*}{}                     & \multirow{2}{*}{\parbox{2cm}{\centering Configuration \\ number}} & \multicolumn{3}{c|}{Number of frame}                                  & \multirow{2}{*}{\parbox{2cm}{\centering Multi-scale \\ feature}} & \multirow{2}{*}{\parbox{2cm}{\centering Recurrent unit}} & \multirow{2}{*}{MC}  & \multirow{2}{*}{$\mathcal{L}_{pre}$} & \multirow{2}{*}{$\mathcal{L}_{recur}$} & \multirow{2}{*}{$\mathcal{L}_{hf}$} & \multirow{2}{*}{PSNR$\uparrow$} & \multirow{2}{*}{SSIM$\uparrow$} \\ \cline{3-5}
                                      &                                   & 3                     & 5                     & 7                     &                                      &                                 &                       &                                         &                                           &                                       &                                              &                                              \\ \hline\hline
Low-dose                              & (0)                               &                       &                       &                       &                                      &                                 &                       &                                         &                                           &                                       & 36.0914                                      & 0.9353                                       \\ \specialrule{0.6pt}{0pt}{0pt}
\multirow{6}{*}{\parbox{2cm}{\centering First \\ training step}}  & (1)                               &                       & \checkmark &                       & \checkmark                & \checkmark           &                       &                                         &                                           &                                       & \underline{37.9258}                                & \underline{0.9756}                                 \\ \cline{2-13} 
                                      & (1-1)                             & \checkmark &                       &                       & \checkmark                & \checkmark           &                       &                                         &                                           &                                       & 37.1097                                      & 0.9739                                       \\ \cline{2-13} 
                                      & (1-2)                             &                       &                       & \checkmark & \checkmark                & \checkmark           &                       &                                         &                                           &                                       & 37.4371                                      & 0.9743                                       \\ \cline{2-13} 
                                      & (1-3)                             &                       & \checkmark &                       & - (Without)                              & \checkmark           &                       &                                         &                                           &                                       & 37.4943                                      & 0.9727                                       \\ \cline{2-13} 
                                      & (1-4)                             &                       & \checkmark &                       & - (Only 3$\times$3)                               & \checkmark           &                       &                                         &                                           &                                       & 37.7203                                      & 0.9738                                       \\ \cline{2-13} 
                                      & (1-5)                             &                       & \checkmark &                       & \checkmark                &  -                               &                       &                                         &                                           &                                       & 37.6038                                      & 0.9740                                       \\ \specialrule{0.6pt}{0pt}{0pt}
\multirow{4}{*}{\parbox{2cm}{\centering Second \\ training step}} & (2)                               &                       & \checkmark &                       & \checkmark                & \checkmark           & & \checkmark                   &                                           &                                       & 38.1624                                      & 0.9785                                       \\ \cline{2-13} 
                                      & (3)                               &                       & \checkmark &                       & \checkmark                & \checkmark           & \checkmark & \checkmark                   & \checkmark                     &                                       & 38.5765                                      & 0.9799                                       \\ \cline{2-13} 
                                      & (4)                               &                       & \checkmark &                       & \checkmark                & \checkmark           & \checkmark & \checkmark                   & \checkmark                     & \checkmark                 & \textbf{39.1201}                             & \textbf{0.9803}                              \\ \cline{2-13} 
                                      & (4-1)                             &                       & \checkmark &                       & \checkmark                & \checkmark           &  -                     & \checkmark                   & \checkmark                     & \checkmark                 & 38.8043                                      & 0.9800                                       \\ 
\bottomrule[1pt]
\end{tabular}%
}
\vspace{-0.18cm}

\end{table*}

\begin{figure*}[ht]
\centering
\includegraphics[width=\linewidth]{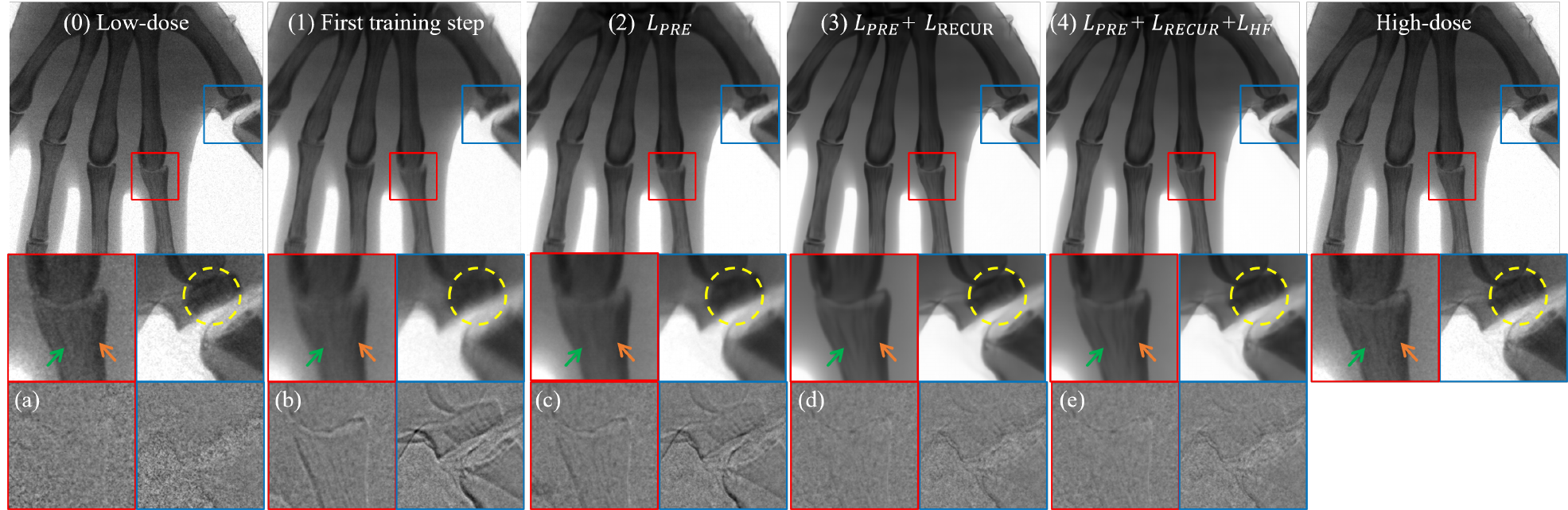}
\caption{Qualitative denoised results from the ablation study corresponding to individual model configurations (e.g., (0)-(4)) listed in Table~\ref{Table2}. Top row: Denoised images. Middle row: Enlarged sections of the ROI in red and blue rectangles in the top row, first image. Bottom row: (a-e) Difference images between denoised results produced by each ablation study model configuration ((0)-(4)) and the ground-truth high-dose images. The Green and orange arrows, along with the dashed yellow lines, highlight the details of fine tissue structures.}
\label{Fig8}
\end{figure*}

We conducted an ablation study to assess the individual contributions of components in the proposed unsupervised denoising framework. We measured the influence of each component using standard quantitative metrics. The PSNR and SSIM scores for various component configurations (1-4) are presented in Table~\ref{Table2}, with their visual comparisons displayed in the corresponding Fig.~\ref{Fig8}. The last row of Fig.~\ref{Fig8} highlights the difference images between the denoised outputs and the high-dose reference images.

\subsubsection{Effectiveness of the Multi-frame input, multi-scale feature extraction, and recurrent unit} In the initial phase of our study, we explored the design choices of the proposed method through an ablation study, examining the individual contributions of its components. The quantitative results are presented in Table~\ref{Table2}. Initially, we established a baseline by setting the number of multi-frame inputs to five (Configuration 1), observing that neither three nor seven frames significantly improved performance (Configuration 1 vs. 1-1 or 1 vs. 1-2). Furthermore, employing solely 3$\times$3 convolutional filters resulted in better PSNR and SSIM metrics compared to configurations without multi-scale feature extraction (Table~\ref{Table2}; Configuration 1 vs. 1-3), and the use of 3$\times$3, 5$\times$5, and 7$\times$7 filters provided a progressive enhancement in both PSNR and SSIM (Table~\ref{Table2}; Configuration 1 vs. 1-4). Additionally, the absence of the recurrent unit resulted in a decrease in PSNR (Table~\ref{Table2}; Configuration 1 vs. 1-5), demonstrating that each component significantly contributes to the overall denoising performance.

\subsubsection{Effectiveness of the Second Training Step and PRE loss \texorpdfstring{$\mathcal{L}_{\text{pre}}$}{L\_PRE}}
A single-stage unsupervised training approach for MSR2AU-Net demonstrated noise reduction capabilities, as evidenced by the increased PSNR and SSIM values (Table~\ref{Table2}; Configuration 1 vs. 0). However, it was ineffective in mitigating uncorrelated structured noise induced by motion, leading to significant edge loss, as visible in the difference images in the last row of Fig.~\ref{Fig8}. By introducing a two-step training strategy and incorporating the PRE loss (Configuration 2), we observed improved quantitative metrics (Table~\ref{Table2}; Configuration 2 vs. 1) and visually confirmed the effective restoration of edges compromised by motion (Fig.~\ref{Fig8}; Configurations 2 vs. 1). These findings validate the efficacy of the proposed two-step training strategy and incorporating the PRE loss (Configuration 2) in enhancing the fluoroscopy image quality.

\subsubsection{Effectiveness of the RECUR loss \texorpdfstring{$\mathcal{L}_{\text{recur}}$}{L\_RECUR}} In this analysis, we evaluated the influence of incorporating an additional recursive filter-based loss, denoted as $\mathcal{L}_{recur}$, on the denoising performance. In Table~\ref{Table2} (Configuration 2 vs. 3), integrating the RECUR loss $\mathcal{L}_{recur}$ contributes to a notable enhancement in PSNR values. Visually, when compared to Fig.~\ref{Fig8} (Configuration 2 vs. 3), the inclusion of the RECUR loss $\mathcal{L}_{recur}$ results in finer and more distinct structures, as indicated by the green and orange arrows and circular dashed yellow lines, signifying an improvement in detail preservation.

\subsubsection{Effectiveness of the HF loss \texorpdfstring{$\mathcal{L}_{\text{hf}}$}{L\_HF} and the Motion Compensation (MC)} The introduction of the high-frequency loss $\mathcal{L}_{hf}$, to accentuate high-frequency details by learning discrepancies through a cross-fusion matrix, is a significant enhancement. The integration of the high-frequency loss $\mathcal{L}_{hf}$, as evidenced in Table~\ref{Table2} (Configuration 4), culminates in the highest PSNR and SSIM metrics, underscoring its efficacy in bolstering denoising performance. Furthermore, it yields sharper delineation of the fine structures, as demarcated by the circular dashed yellow lines in Fig.~\ref{Fig8} (Configuration 4), underscoring its contribution to image clarity. The removal of the motion compensation module results in decreased PSNR and SSIM metrics in the second training step compared to Configuration 4, suggesting that motion compensation plays a critical role in achieving satisfactory denoising performance.

\section{Discussion}

This work addresses the pivotal challenge of noise reduction in low-dose fluoroscopy images, particularly those affected by motion artifacts. The proposed unsupervised approach offers a significant advantage by reducing the reliance on clean target data, which is often scarce or challenging to obtain in medical imaging. By effectively targeting both uncorrelated and correlated noise components through the integration of knowledge distillation, motion compensation, and recursive filtering techniques, the proposed method ensures comprehensive noise reduction while preserving critical edge details necessary for accurate medical diagnoses.

When compared to similar works, such as UDDN~\cite{luo2020ultra} and the hybrid 3D/2D-based deep CNN~\cite{zhang2019hybrid}, our method demonstrates several key improvements. UDDN~\cite{luo2020ultra} leveraged dense connections within CNN frameworks for efficient information reuse, and the hybrid 3D/2D-based deep CNN~\cite{zhang2019hybrid} utilized a combination of 2D and 3D data for enhanced noise reduction in X-ray angiography. However, these methods depend heavily on supervised learning, requiring large datasets of paired noisy and clean images—a requirement that is difficult to meet in clinical settings due to issues like patient movement and the risks associated with increased radiation exposure.

In contrast, our approach successfully implements an unsupervised two-step training framework that not only reduces noise effectively but also preserves the fine details of fluoroscopic images. The incorporation of a high-frequency loss component further enhances the preservation of texture details, setting our method apart from existing techniques that may struggle with issues like over-smoothing or motion blurring. This approach also demonstrates that unsupervised learning methods can approach, and even match, the performance benchmarks set by state-of-the-art supervised methods, such as MCDnCNN~\cite{wu2020combined} and FastDVDNet~\cite{tassano2020fastdvdnet}. While these supervised models achieved notable results in noise reduction, our unsupervised method outperforms them in terms of edge preservation and overall noise reduction, as evidenced by superior PSNR and SSIM values in our experiments.

One noteworthy observation from this study is the convergence of unsupervised learning methods toward the performance levels typically seen in supervised counterparts. This convergence underscores the advancements in unsupervised learning techniques and highlights their potential for broader application in clinical practices, where obtaining annotated data remains challenging.

Moreover, the study results demonstrate the successful application of the proposed denoising framework to sequential low-dose fluoroscopy images, which present unique challenges due to their semantically sparse information and motion-induced artifacts. The utility of this approach is not confined to image denoising; it can be readily extended to other tasks of restoration and enhancement across various sequential images or video frames, particularly those characterized by object motion or limited semantic information.

The versatility of the proposed method is further highlighted by its successful extension beyond fluoroscopy to other imaging modalities, such as low-dose CT. In conventional approaches, low-dose CT slices are often processed independently; however, by treating sequential slices along the channel dimension similarly to temporal video frames, the anatomical variations observed between adjacent slices can be considered as correlated noise (i.e., $\delta_{i}$ in Eq.~\eqref{eq:delta-correlated-noise}). This similarity between 3D spatial CT data and 3D temporal video data underscores the adaptability of our method, which has been effectively applied to low-dose CT denoising tasks as well as other static volumetric imaging modalities like MRI.

This adaptability and the demonstrated performance gains suggest that the proposed method could have a broad impact on various medical imaging contexts, offering a scalable solution to the pervasive problem of noise in low-dose imaging.

\section{Conclusion}
In this work, we presented an unsupervised denoising framework for low-dose fluoroscopy imaging that effectively reduces noise while preserving critical image details. The method’s ability to perform well without relying on clean target data positions it as a valuable tool in medical imaging, where such data are often difficult to obtain.
Our framework, which integrates knowledge distillation, motion compensation, and recursive filtering, demonstrates that unsupervised methods can achieve performance comparable to that of supervised approaches. The proposed method is not only effective for fluoroscopy but also highly adaptable, offering potential applications in both dynamic video imaging modalities and static volumetric imaging modalities, such as CT and MRI, demonstrating its broad scalability.
Future work should focus on optimizing the framework for real-time applications and further validating its performance in clinical settings to fully realize its potential in improving diagnostic accuracy.

\bibliographystyle{IEEEtran}
\bibliography{IEEEabrv,refs}

\end{document}